\documentclass[showpacs,preprintnumbers,amsmath,amssymb,superscriptaddress]{revtex4}

\usepackage{graphicx}
\usepackage{dcolumn}
\usepackage{bm}
\usepackage{epsfig}

\begin{document}

\title{Spectral line shape modeling and ion temperature fluctuations in tokamak edge plasmas}

\author{Y. Marandet}
\affiliation{Department of Physics, University of Florida, P.O. Box
118440, Gainesville, Florida 32611-8440,USA}
\author{J. W. Dufty}
\affiliation{Department of Physics, University of Florida, P.O. Box
118440, Gainesville, Florida 32611-8440,USA}

\begin{abstract}
In this work, we use a passive advection model for ion temperature
fluctuations, in order to investigate their effects on Doppler
Spectral line shapes. The relevance of the model is discussed in the
framework of the Braginskii equations, and the subsequent
Probability Density Function evaluation relies on results obtained
in neutral fluids. The resulting Doppler line profiles are shown to
exhibit characteristic exponential tails.
\end{abstract}


\maketitle

\section{Introduction}

The understanding of turbulence in fusion devices has made great
progress since it was first realized that turbulent fluctuations
could explain the anomalously high level of cross magnetic field
transport which was plaguing experiments. Theoretical studies, many
of which have focused on the
identification of underlying linear or non linear instabilities \cite%
{Horton1999}, have produced valuable results. Direct Numerical
Simulation, which is the only tool able to encompass the
complexity of the problem (non linear equations, multiple fields
and complex geometry), has shed additional
light on the respective contributions of these instabilities (e.g. \cite%
{Scott2005}). Edge plasma turbulence has specific properties and is,
in particular, characterized by high fluctuation levels which can
rise up to 30 \% \cite{Bretz1997,Evensen1998}. The development of
sophisticated diagnostics methods now provides detailed experimental
data on density fluctuations (e. g. \cite{Huber2005}), which impose
more stringent constraints on the theory. However, very few
measurements of ion temperature fluctuations have been reported to
date. A notable exception is Ref. \cite{Evensen1998}, in which High
Frequency Charge Exchange Recombination Spectroscopy is used to show
that ion temperature fluctuations rates can be large. In previous
works \cite{Marandet2004,Marandet2005}, we have investigated the
possible role of turbulence on Doppler line shapes obtained by
passive spectroscopy. Ion temperature fluctuations were shown to
lead to significant modifications in the line wings, provided their
PDF (Probability Density Function) had fat tails. This suggests that
passive spectroscopy, for which the experimental set-up is
relatively simple, might provide information on ion temperature
fluctuations. Here, making use of results obtained in neutral
fluids, we show that even a very simplified model for ion
temperature fluctuations, i.e. passive advection, leads to non
Gaussian PDFs. The corresponding asymptotic behavior of the Doppler
line profiles is obtained analytically.

\section{Doppler spectra in turbulent plasmas}

The spectra $I_{m}(\Delta \lambda )$ measured using passive
spectroscopy is obtained from integration both during the
acquisition time of the spectrometer $\tau$ and along the line of
sight (LOS) $(Oz)$. The corresponding expression is

\begin{equation}
I_{m}(\Delta\lambda)=\frac{1}{\tau}\int_{0}^{\tau}
dt\frac{1}{L}\int dz B(z,t) I(\Delta\lambda,z,t),
\end{equation}

\noindent where $L$ is the emitting zone length, $B(z,t)$ the
brightness of the line and $I(\Delta \lambda ,z,t)$ the local line
shape. Both of these quantities are determined by the plasma
parameters and the line under study. Here we consider $D_{\alpha}$
($n=3$ to $n=2$ transition, $\lambda_{0}=6561$ \AA), which is
intense and routinely monitored. The brightness essentially depends
on the electron density and temperature. For ionizing edge plasma
conditions, that is $N_{e}=10^{19}-10^{20}$ m$^{-3}$ and
$T_{e}=10-100$ eV, Stark effects can be safely neglected. We will
therefore deal with the Doppler line profile of one Zeeman
component. Here we only consider the contribution from the class of
neutrals created locally by charge exchange, $I(\Delta \lambda
,z,t)\rightarrow $ $I^{cx}(\Delta \lambda ,\mathbf{X})$. In fact,
neutrals created via other channels, i.e molecular dissociation and
charge exchange in the inner plasma are not strongly coupled to the
ions at the edge \cite{Hey2004}, i.e. to edge turbulence. These
neutrals can be accounted for using neutral transport codes. The
charge exchange neutrals Velocity Distribution Function (VDF) along
the LOS, $f^{cx}(v,\mathbf{X})$, is related to the Doppler profile
by $f^{cx}(v,\mathbf{X})dv=I^{cx}(\Delta \lambda ,\mathbf{X})d\Delta
\lambda$. Here it has been made explicit that the space and time
dependence of the spectra and VDF occurs only
through the local ion fluid fields $%
\mathbf{X}(z,t)$ (local number and charge densities, species
temperatures, and flow velocities), since the ion VDF is assumed to
be a local Maxwellian. We consider the case where the LOS is such
that the average gradients of these fields along its
direction (e.g., parallel to the magnetic field lines) are weak . It was shown in Ref. %
\cite{Marandet2004,Marandet2005} that the measured spectra can then
be written as

\begin{equation}  \label{eq:Dp}
I^{cx}_{m}(\Delta\lambda)=\int d\mathbf{X}W(\mathbf{X})B(\mathbf{X}%
)I^{cx}(\Delta\lambda,\mathbf{X}),
\end{equation}

\noindent where $W(\mathbf{X})$ is the joint PDF of the fields. To
proceed, the role of the different fluid fields can be investigated
separately. For Doppler line shapes, fluid velocity and temperature
fluctuations give rise to the main effects, because the shape of the
local VDF $f_{cx}$ does not depend on $N_{e}$. We further simplify
the problem by considering ion temperature fluctuations alone. The
rationale behind that assumption is twofold : first, the role of
velocity fluctuations is well known in plasma spectroscopy
\cite{Griem1997}. Secondly, the analysis carried out in our previous
work showed that line wings are significantly affected by
temperature fluctuations characterized by a slowly decaying tail.
Therefore, we expect that inclusion of velocity fluctuations would
not significatively shroud the conclusions drawn here, since line
wings could only be affected by fluid velocities several times
larger than the ion thermal velocity. Moreover, if the LOS is
parallel to the magnetic fields, there are conditions under which
parallel velocity fluctuations are small while temperature
fluctuations are not \cite{Evensen1998}. As a result, the remainder
of this contribution will be devoted to temperature fluctuations.



\section{Simplified model for the ion temperature}

\label{sec:tempmodel}

We start from the Braginskii equations \cite{Braginskii1965}, and
focus on the ion thermal balance

\begin{equation}
\frac{3}{2} d_{t}(n_{i}T_{i})+\frac{5}{2}n_{i}T_{i}\nabla\cdot\mathbf{v}%
_{i}+\pi :: \sigma +
\nabla\cdot\mathbf{q}_{i}=Q_{i}(T_{i},\textbf{v}),
\end{equation}

\noindent which we simplify using the drift wave ordering. Here $%
d_{t}=\partial _{t}+\mathbf{v}_{i}\cdot \nabla $ stands for the
convective derivative, $\mathbf{q}_{i}$ for the thermal flux, $\pi
_{\alpha \beta }$ is the stress tensor, $\sigma _{\alpha \beta
}=\partial _{\alpha }v_{\beta }$ and $Q_{i}$ is the energy exchange
with other species. We consider fluctuations having a small scale
$\tilde{l}$ with respect to the quasi equilibrium background of
radial gradient length $L_{\perp }$. Typically $\rho _{s}\lesssim
\tilde{l}$, where $\rho _{s}$ is the ion Larmor radius calculated
with the electron temperature. For each field $X$, we distinguish
between its time averaged part $\bar{X}$ and its fluctuating part $%
\widetilde{X}$. The small scale parameter with respect to which the
fluid equations are expended is $\delta =\tilde{l} /L_{\perp }$. The
perpendicular velocity is given by $\mathbf{v}_{i\perp }=\mathbf{v}_{E}+\mathbf{v}%
_{i}^{\star }+\mathbf{v}_{i}^{p}$, that is by the sum of the
electric drift, ion diamagnetic and ion polarization velocities. The
diamagnetic velocity does not directly advect temperature, because
for low $\beta $ (the ratio of the kinetic to the magnetic pressure)
its effect cancels out with that of the diamagnetic thermal flux, up
to magnetic curvature terms which
will be neglected in the following. The same approximations ensure that $%
\nabla \cdot \widetilde{\mathbf{v}}_{E}\simeq 0$, i.e. the
fluctuating electric drift velocity flow is incompressible. The ion
polarization velocity is of higher order in $\delta $ and we further
neglect parallel advection, which is justified in the drift ordering
\cite{Scott1997}. The resulting equation for ion temperature
fluctuations is then

\begin{equation}  \label{eq:passivescal}
\left[\partial_{t}+(\overline{\mathbf{v}}_{E}+\widetilde{\mathbf{v}}_{E})\cdot\nabla-\kappa_{i\perp}%
\nabla^{2}\right]\widetilde{T}_{i}=\widetilde{S}(\mathbf{r},t),
\end{equation}

\noindent where $\kappa_{i\perp}$ is the ion perpendicular thermal
diffusivity, and all the terms on the r.h.s. have been lumped into a
source, that is
$\widetilde{S}(\mathbf{r},t)=\widetilde{\mathbf{v}}_{E}\cdot \nabla
\bar{T_{i}}-\overline{\widetilde{\mathbf{v}}_{E}\cdot \nabla
\widetilde{T_{i}}}+\widetilde{Q}_{i}$. The energy exchanges between
electrons and ions $Q_{ie}$ and ions and neutrals $Q_{in}$ are
neglected on the turbulent time scales, i.e.
$\widetilde{Q}_{i}=Q_{i}-\bar{Q}_{i}\simeq 0$. Also, the second term
is zero by homogeneity, so that
$\widetilde{S}(\mathbf{r},t)=\widetilde{\mathbf{v}}_{E}\cdot \nabla
\bar{T_{i}}$. The mean temperature gradient therefore naturally
provides a source term for the fluctuations through its coupling to
$\widetilde{\mathbf{v}}_{E}$. Since the scale of the turbulent
fluctuations is such that $\tilde{l}\ll L_{\perp }$, in the
following the gradient will be assumed to be constant and directed
along $x$, namely $\nabla_{x}\bar{T_{i}}=g$. Advection by the mean
flow $\overline{\mathbf{v}}_{E}$ in Eq. (\ref{eq:passivescal}) can
be formally removed by introducing suitable Lagrangian coordinates.
This does not interfere with the following developments, so that we
will assume $\overline{\mathbf{v}}_{E}=0$. Rigorously speaking, the
evolution of $\widetilde{\mathbf{v}}_{E}$ is not independent from
that of $\widetilde{T}_{i}$ (see for example the system of equations
considered in Ref. \cite{Scott2005}). However, the ion temperature
does not play any role in the parallel electron dynamics, in
contrast to the electron temperature which is strongly coupled to
the electrostatic potential, and hence to
$\widetilde{\mathbf{v}}_{E}$. In addition, many if not most edge
turbulence studies carried out so far did not include ion
temperature fluctuations, by considering cold ions ($T_{i}\ll
T_{e}$). This approximation is justified partly by the fact that ion
temperature does not play a fundamental role for the instabilities
which are thought to be dominant in edge plasmas (collisional drift
waves, resistive ballooning). In fact, inclusion of
$\widetilde{T}_{i}$ does not qualitatively change the nature of the
turbulence observed in the simulations \cite{Scott2005}, and the few
experimental results available are not inconsistent with passive
advection, as pointed out in \cite{Evensen1998}. In this work, we
therefore treat ion temperature as a passive scalar driven by the
fluctuating electric drift velocity field. This has the major
advantage to lead to an analytically tractable model. Let us now
specify the typical values of the P\'{e}clet and the Prandtl
numbers, which will play a role in the next section, for edge plasma
turbulence. The P\'{e}clet number, $\textrm{Pe}=\tilde{l}
\tilde{v}/\kappa_{i\perp} $, where $\tilde{l}$ and $\tilde{v}$ are
the typical length scales and velocity of the turbulent
fluctuations, controls the relative importance of advection and
diffusion. If we take $\tilde{l}\sim 10\rho _{s}\sim 2\times
10^{-3}$ m and $\tilde{v}\sim \tilde{v}_{E}\sim
\tilde{v}_{i}^{\star}\sim T_{i}/(eBL_{\perp })\sim 75$ m.s$^{-1}$
with $L_{\perp }=2$ cm, we get $\text{Pe}\sim 10^{2}\gg 1$, so that
advection dominates the heat transport. The Prandtl number
$\textrm{Pr}=\nu_{i\perp} /\kappa_{i\perp}$, where $\nu_{i\perp}$ is
the perpendicular ion viscosity is of order unity since
$\nu_{i\perp} /\kappa_{i\perp} \sim 1$ \cite{Braginskii1965}.


%
%




\section{Review of analytical calculations of the passive scalar PDF}

Since $\tilde{T}_{i}$ and $\tilde{\textbf{v}}$ are rapidly
fluctuating quantities, we adopt a stochastic description of the
velocity field. For the sake of tractability, the latter is
described by a Gaussian probability density \textit{functional},
whose correlation function is chosen so as to reproduce some of the
main features of turbulence \cite{Schraiman1994}, that is
\begin{equation}\label{eq:corr}
    C_{\alpha\beta}(\mathbf{r},t)=\langle v_{\alpha}(\mathbf{r},t)v_{\beta}(0,0)\rangle=\Pi_{\alpha\beta}V\xi\exp(-|\mathbf{r}|/\xi)\delta(t)
\end{equation}
where $\Pi_{\alpha\beta}(\mathbf{r})$ ensures consistency with
incompressibility \cite{Falkovitch2001} (at zeroth order in
$\mathbf{r}$, $\Pi_{ab}=\delta_{ab}$), and the white noise limit has
been taken. Here, the correlation length $\xi$ and the typical
velocity $V$ are such that $\xi\sim \tilde{l}$ and $V\sim\tilde{v}$.
The time average is thus replaced by an ensemble average denoted by
$\langle\cdot\rangle$. In this section, we show how the ion
temperature PDF $W(\theta)=\langle\delta (\theta
-\widetilde{T}_{i}(\mathbf{r},t))\rangle$, or equivalently its
Fourier transform $Z(\lambda)=\langle\exp (-i\lambda \widetilde{T}_{i}(\mathbf{r}%
,t))\rangle$, can be obtained from Eq. (\ref{eq:passivescal}). Here,
$\theta$ is a sample space variable, and $\lambda$ its conjugated
Fourier variable. Mathematically, the brackets denote \textit{functional} integration over the field $\widetilde{%
\mathbf{v}}$ at all space and time points. The PDF does not depend
on $t$ and $\mathbf{r}$ because of stationarity and homogeneity, the
latter stemming from the fact that the average temperature gradient
is constant. The derivation presented here is drawn from the neutral
fluid community, which has given much attention to the passive
advection problem. Indeed, its study allows to unravel important
effects also at play in Navier-Stokes turbulence. The passive scalar
equation is linear in $\widetilde{T}_{i}$, but bilinear in the
fluctuations and can thus generate non linear effects. The scalar
PDF $W(\theta)$ can therefore have strong non Gaussian features even
though the velocity field is Gaussian. This is very well illustrated
by the early work of Sinai and Yakhot, who found power law tails for
the PDF in homogeneous decaying turbulence \cite{Sinai1989}.  The
derivation relies on a transport equation for the PDF, which
involves coefficients whose temperature dependence has to be
determined from rather uncontrolled assumptions. This difficulty
arises because diffusion introduces correlations between the scalar
field and its gradients. The physical reason for this is most easily
understood using a Lagrangian
picture. In the weak thermal diffusion limit (i.e. large P\'{e}clet number $%
\text{Pe}\gg 1$, see below), advection is the dominant initial
effect and
the fluid particles follow the Lagrangian trajectories defined by $d_{t}%
\mathbf{R}(t)=\mathbf{v}(\mathbf{R}(t),t|\mathbf{R_{0}})$ where $\mathbf{R}%
(t=0)=\mathbf{R}_{0}=\mathbf{r}$. The velocity field is
linearized around the Lagrangian trajectory $\left( \widetilde{v}_{\alpha }(\mathbf{r}%
,t)\rightarrow \widetilde{v}_{\alpha }(\mathbf{R}(t),t)+\sigma
_{\alpha \beta }(\mathbf{R}(t),t)\left( r_{\beta }-R_{\beta
}(t)\right) \right) $,
where $\sigma _{\alpha \beta }(\mathbf{R}(t),t)=\partial _{\alpha }v_{\beta }(\mathbf{R}%
(t),t)$ is the strain field along the trajectory. Two points in a
fluid element, initially close will be displaced by this strain
field. Due to the condition of incompressibility, one direction will
involve an exponentially large separation as a function of time. As
a consequence of this so-called "advective stretching", the spectra
of temperature fluctuations are extended to smaller and smaller
scales. When the diffusive scale is eventually reached, regions of
large gradient are selectively dissipated, introducing correlations
between the temperature field and its gradient \cite{Kimura1993}.
The PDF obtained in Ref. \cite{Sinai1989} has been used as an
example in our previous work \cite{Marandet2004,Marandet2005}, but
from the section 3 it is clear that an additive noise should also be
taken into account. The corresponding
 transport equation for the PDF could straightforwardly be written down,
but finding its solution would require further assumptions on the
correlations between $\widetilde{T}_{i}$ and $\widetilde{S}$
\cite{Pierrehumbert2000}. From the discussion above it is obvious
that a Lagrangian description would allow to capture more of the
physics. The remainder of this section will be devoted to show how
this unfolds. The first difficulty when carrying out the average in
$Z(\lambda)$ lies in the fact that there are two different sources
of noise associated with $\widetilde{\textbf{v}}_{E}$, one additive
($\widetilde{S}=g\widetilde{v}_{Ex}$) and one multiplicative
($\widetilde{\textbf{v}}_{E}\cdot\nabla$). An elegant way to
disentangle their respective contributions is to calculate the
Green's function $G$ for Eq. (\ref{eq:passivescal}), satisfying
\begin{equation}
\widetilde{T}_{i}(\mathbf{r},t)=\int d\mathbf{r}^{\prime }dt^{\prime }G(\mathbf{r%
},t|\mathbf{r}^{\prime },t^{\prime
})g\widetilde{v}_{Ex}(\mathbf{r}^{\prime },t^{\prime }).
\end{equation}

\noindent This has the advantage to allow averaging on the additive
noise as a first step. The latter is
independent of the statistics of $\widetilde{\textbf{v}}_{E}$ in $G(\mathbf{r},t|\mathbf{r%
}^{\prime },t^{\prime })$ in the limit of white noise. Physically
speaking this occurs when the time scale for velocity field
correlations is short compared to other relevant time scales for
$\widetilde{T}_{i}$. In the following, all results quoted are in
this limit.  The resulting expression for the characteristic
function is

\begin{equation}
Z(\lambda )=\left\langle \exp \left( -\lambda ^{2}\int_{-\infty }^{0}dt\int d%
\mathbf{k}\widehat{D}(\mathbf{k})|\widehat{G}(\mathbf{k},t)|^{2}\right)
\right\rangle _{M},
\end{equation}

\noindent where $\langle\cdot\rangle_{M}$ stand for the ensemble
average over the multiplicative noise, i.e. the $\textbf{v}$
dependance of the Green function. $\widehat{D}(\mathbf{k})$
is the Fourier transform of the additive noise correlation function $D(%
\mathbf{r})=\langle
\widetilde{S}(0,t)\widetilde{S}(\mathbf{r},t)\rangle_{A}=g^{2}C_{xx}(\mathbf{r})
$. This approach was first carried out in Ref. \cite{Schraiman1994}
using a path integral formulation to calculate the Green's function,
assuming the above linearization of the velocity field around the
Lagrangian trajectory. Strictly speaking, the latter is valid for
$\textrm{Pr}\gg 1$, but is thought to be also relevant to the cases
where $\textrm{Pr}\sim 1$ \cite{Schraiman1994}. The same
approximation allows to expand $C_{xx}(\mathbf{r})$ in terms of
$|\mathbf{r}|/\xi$. The Green's function, Fourier transformed with
respect to the starting point, is then found to be

\begin{equation}
\widehat{G}(0,0|\mathbf{k}_{0},t)=e^{i\mathbf{k}_{0}\cdot \mathbf{r}%
(t;0,0)}\exp \left( -\frac{\kappa_{i\perp} }{2}\int_{t}^{0}dt^{\prime }\mathbf{k}%
^{2}(t^{\prime })\right) ,  \label{eq:Greens}
\end{equation}

\noindent where the wave number $\mathbf{k}(t)$ is defined by the
time ordered product ($\mathcal{T}$ stands for latest times on left)

\begin{equation}\label{eq:kt}
k_{\alpha}(t)=\mathcal{T}\left[%
\exp\left(-\int_{t}^{0}\sigma(t^{\prime})dt^{\prime}%
\right)\right]_{\alpha\beta}k_{0\beta}.
\end{equation}

\noindent Here $\mathbf{r}(t;0,0)$ is the backwards Lagrangian
trajectory, i.e. the position at which the fluid particle which is
at time $t=0$ at position $\mathbf{r}=0$ was at time $t<0$, not
taking into account diffusion. The second term in Eq.
(\ref{eq:Greens}) is a damping term originating from diffusion,
where the time dependence of $\mathbf{k}(t)$ describes the effect
of stretching integrated along the Lagrangian trajectory. The norm
of $\mathbf{k}$ grows exponentially for the velocity field
considered here, so that this second term basically behaves as a
step
function at a time $t^{\star }$, which can be defined as $|\widehat{G}(0,0|\mathbf{k}%
_{0},t^{\star })|\sim 1/2$. The larger the strain along the
trajectory, the smaller $t^{\star }$. For $t<t^{\star }$ the fluid
particle follow the diffusionless Lagrangian trajectory with
constant temperature, and then the latter is quickly homogenized
because of the large gradients induced by stretching.  Since the
modulus of the Green's function only depends on the velocity field
through the strain tensor $\sigma $ (see Eq.
(\ref{eq:Greens})-(\ref{eq:kt})), the velocity average is then
converted into an average over realizations of $\sigma $. Carrying
out the calculation is tedious, but only involves some additional
mild approximations \cite{Schraiman1994}. The resulting PDF has a
Gaussian core and an asymptotical behavior which is exponential,
that is using non dimensional quantities

\begin{equation}
W(\theta)\sim \frac{1}{|\theta|^{1/2}}e^{-\gamma|\theta|},
\end{equation}

\noindent where $\gamma=(d+6)/2d$ is the secular growth rate, and
$d$ the dimensionality of the problem ($\gamma=2$ in 2D)
\cite{Schraiman1994}. The same result has been obtained using
Jensen's field theoretical formalism \cite {Falkovich1996}. Taking
an additive noise into account, as required by the derivation of Eq.
(\ref{eq:passivescal}), therefore leads to a PDF which has an
exponential tail instead of the power law tail obtained for decaying
turbulence. The PDF is however still markedly non Gaussian.

\section{Calculations of Doppler spectra}

\begin{figure}[htb]
\includegraphics[width=.45\textwidth]{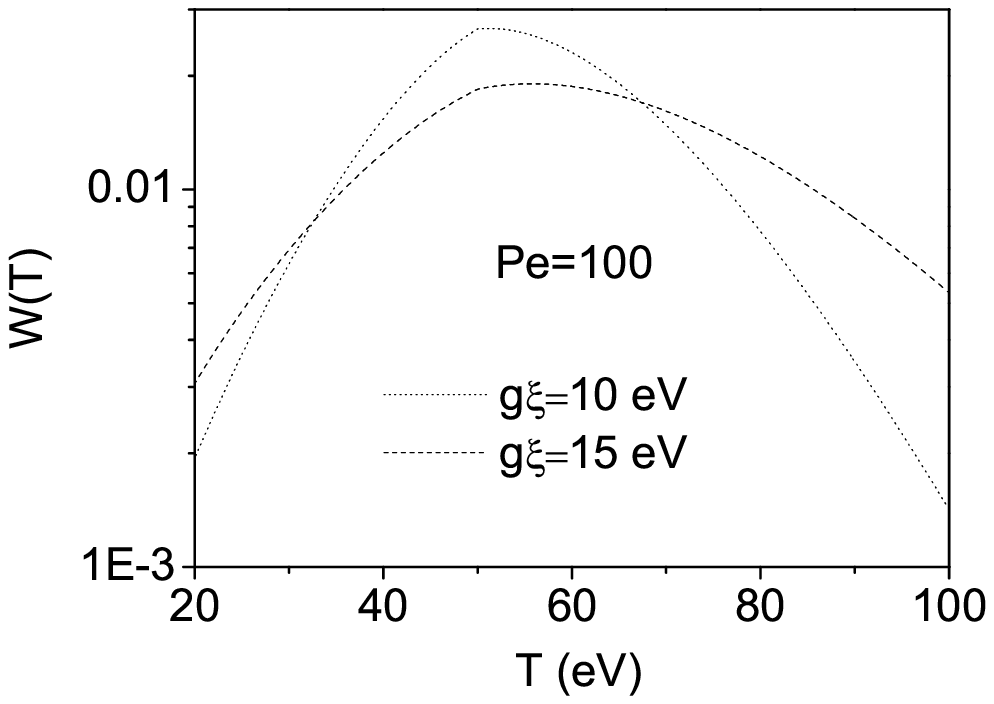}~a)
\hfil
\includegraphics[width=.45\textwidth]{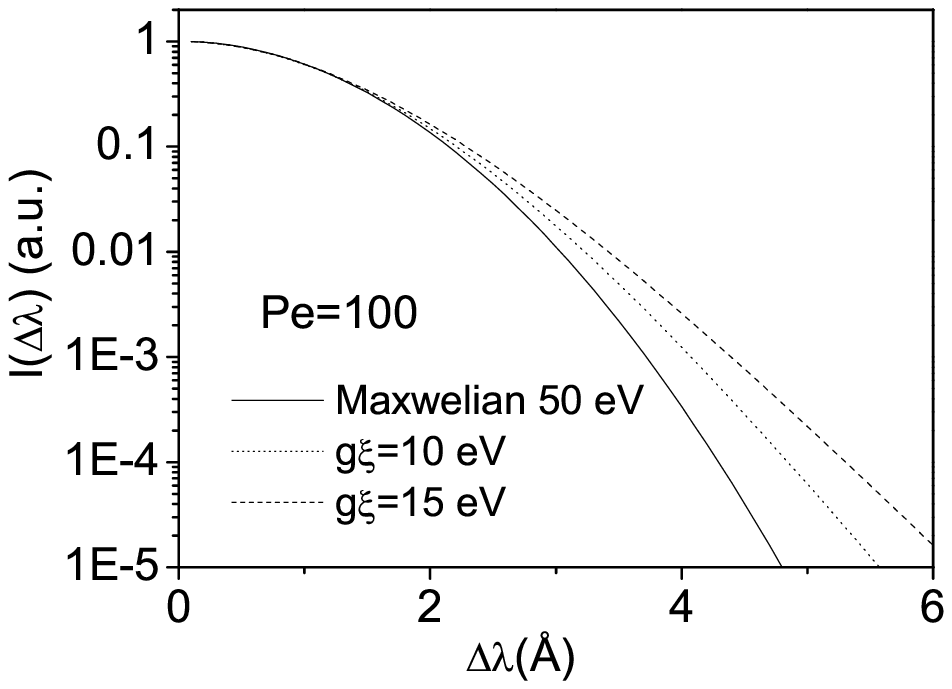}~b)
\caption{a) Plot of the PDF of Eq. (\ref{eq:schraimanPDF}) for
$T_{0}=50$ eV, $g\xi=10$ eV, $\textrm{Pe=10}^{2}$,
$V\xi/\mathcal{K}=1$ (dotted line), and $g\xi=15$ eV (dashed line) .
b) Plot of the resulting Doppler profile on a logarithmic scale
(solid and dashed lines), showing the asymptotic exponential
behavior. The solid line corresponds to the Gaussian Doppler profile
obtained for 50 eV. \label{fig:1}}
\end{figure}

We now consider the modifications resulting from temperature
fluctuations on line shapes, given the PDF obtained in the previous
section. The Doppler
line shape $I_{m}(\Delta \lambda)$ can be calculated numerically from Eq. (\ref%
{eq:Dp}) for any given $W(\theta )$, noting that the brightness of the line $%
B$ does not sensitively depend on the ion temperature. However, it
is interesting first to obtain an analytical expression of the
profile. This can be done for PDFs behaving asymptotically as
$W(\theta )\sim \theta ^{-\alpha }\exp -\left( \theta /\theta_{0}
\right) ^{\beta }$. The asymptotic behavior of the measured Doppler
profile can then be calculated using a saddle point approximation,
which leads to

\begin{equation}
I_{m}(\Delta\lambda)\propto \frac{1}{|\Delta\lambda|^{\frac{\beta+2\alpha-1}{%
\beta+1}}}\exp\left(-C(\theta_{0},\beta)|\Delta\lambda|^{\frac{2\beta}{\beta+1}%
}\right),
\end{equation}

\noindent where $C(\theta_{0} ,\beta )=(\beta +1)/(\beta \zeta
\theta_{0} )^{\beta /\beta +1}$, with
$\zeta=\lambda_{0}^{2}/(c^{2}m_{i})$, $m_{i}$ being the ion mass and
$c$ the speed of light. For $\beta =0$, the measured profile decays
algebraically in accordance with Ref.
\cite{Marandet2004,Marandet2005}, while for an exponential PDF
($\beta =1$), its decay is exponential too. This implies that ion
temperature fluctuations characterized by an exponential PDF would
lead to a conspicuous behavior in the line wings. As an example, we
consider the following temperature PDF (adapted from Eq. (4.10) in
Ref. \cite{Schraiman1994}), having a Gaussian core and exponential
tails

\begin{equation}\label{eq:schraimanPDF}
    W(\theta)=
    \frac{N}{(g\xi)\sqrt{g\xi(\mathcal{K}/V\xi)\ln \textrm{Pe}}+|\theta|^{3/2}}(\theta+ T_{0})\exp\left(-\frac{2}{g\xi}\frac{\theta^{2}}{g\xi (2\mathcal{K}/V\xi) \ln\textrm{Pe}+\sqrt{2\mathcal{K}/V\xi}|\theta|}\right),
\end{equation}

\noindent where $N$ is a normalization constant, $\xi$ and $V$ are
defined by Eq. (\ref{eq:corr}), and
$\mathcal{K}=\int_{0}^{+\infty}C_{xx}(\mathbf{R}(t),t)dt$ is the
eddy diffusivity, where $\mathbf{R}(0)=0$. The effective Peclet
number $V\xi/\mathcal{K}$ is such that $\mathcal{K}/V\xi=1$
according to Eq. (\ref{eq:corr}). This VDF is plotted on Fig.
\ref{fig:1}a for $T_{0}=50$ eV, $\textrm{Pe}= 10^{2}$, and two
gradient values $g\xi=10$ eV and $15$ eV. The resulting Doppler
profiles are plotted on Fig. \ref{fig:1}b, together with the
Gaussian at $50$ eV corresponding to the fluctuation free case.
The center of the line is not strongly affected, whereas significant
deviations occur in the line wings. The linear dependance
characterizing an exponential fall off in log-linear scale is
clearly seen. These results show that Doppler line profiles would
indeed be affected by ion temperature fluctuations stemming from
passive advection. Comparison to experimental spectra requires
specifically designed measurements with a large dynamic range. With
existing techniques, spectra can be recorded over five orders of
magnitude using CCD detectors operated in specifically optimized
readout mode \cite{CCD1994}. In Ref. \cite{Marandet2003}, we
presented spectra measured in the Tore Supra tokamak, the far line
wings of which were behaving in a way consistent with a power law.
However, reliable determination of the exact nature of the these
deviations (exponential or power law behavior) would require
improved measurements. In fact, the latter spectra were obtained as
a by-product of routine measurements, and therefore line wings were
recorded only over a limited range. These spectra nevertheless
provided clear indications of the presence of an unexpected behavior
in the far line wings, and are a strong incentive for further
investigations.


\section{Conclusions and perspectives}

 We have shown that as a first approximation the ion temperature field
in edge plasmas can be described as a passive scalar, advected by
the fluctuating electric drift velocity field. The average
temperature gradient naturally provides a forcing term for the
fluctuations. The PDF of the passive scalar can be calculated using
results obtained in neutral fluids, and have an exponential fall
off. The latter translates into an exponential fall off for Doppler
line wings. Therefore, passive spectroscopy might provide
information on the statistical properties of ion temperature
fluctuations, about which very few is known at this time. The
effects on Doppler line shapes are observable when the temperature
fluctuation PDF deviates significantly from gaussianity, so that
their study could either uncover, or rule out such behavior. This
work also provides a strong incentive to study the statistical
properties of ion temperature fluctuations by direct numerical
simulation. This would allow to compare the PDF obtained numerically
to those provided by our simplified model.


\begin{acknowledgements}
This work was supported by a collaboration (LRC DSM 99-14) between
the Laboratoire de Physique des Interactions Ioniques et
Mol\'{e}culaires and the D\'{e}partement de Recherches sur la Fusion
Contr\^{o}l\'{e}e, CEA Cadarache, and by the Department of Energy
grant DE FGO2ER54677.
\end{acknowledgements}


\begin{thebibliography}{10}
\bibitem{Horton1999} W. Horton, Rev. Mod. Phys. \textbf{71}, 735 (1999).
\bibitem{Scott2005} B. Scott, Phys. Plasmas \textbf{12}, 062314 (2005).
\bibitem{Bretz1997} N. Bretz, Rev. Sci. Instrum. \textbf{68}, 2927 (1997).
\bibitem{Evensen1998} T. H. Evensen \textit{et al.}, Nucl. Fusion \textbf{38}, 237 (1998).
\bibitem{Huber2005} A. Huber \textit{et al.}, Plasma Phys. Control. Fusion,
\textbf{49}, 409, (2005).
\bibitem{Marandet2004} Y. Marandet \textit{et al.}, Contrib. Plasma Phys. \textbf{44}, 283 (2004).
\bibitem{Marandet2005} Y. Marandet \textit{et al.}, Europhys. Lett. \textbf{69}, 531 (2005).
\bibitem{Hey2004} J. D. Hey \textit{et al.}, J. Phys. B: At. Mol. Opt. Phys.
\textbf{37}, 2543 (2004).
\bibitem{Griem1997} H. R. Griem, Principles of Plasma Spectroscopy, (Cambridge University Press, 1997).
\bibitem{Braginskii1965} S. I. Braginskii, Rev. Plasma Phys.
\textbf{1}, 205 (1965).
\bibitem{Scott1997} B. Scott, Plasma Phys. Control. Fusion,
\textbf{39}, 1635 {1997}.
\bibitem{Schraiman1994} B. I. Schraiman and E. D. Siggia, Phys. Rev.
E \textbf{49}, 2912 (1994).
\bibitem{Falkovitch2001} G. Falkovich \textit{et al.}, Rev. Mod. Phys.
\textbf{73}, 913, (2001).
\bibitem{Sinai1989} Y. G. Sinai and V. Yakhot, Phys. Rev. Lett.
\textbf{63}, 1962 (1989).
\bibitem{Kimura1993} Y. Kimura and R. H. Kraichnan, Phys. Fluids A
\textbf{5}, 2264 (1993).
\bibitem{Pierrehumbert2000} R. T. Pierrehumbert, Chaos \textbf{10},
61 (2000).
\bibitem{Falkovich1996} G. Falkovich \textit{et al.}, Phys. Rev. E
\textbf{54}, 4896 (1996).
\bibitem{CCD1994} J. V. Sweedler \textit{et al.}, Charge Transfer Devices in
spectroscopy, (VCH, 1994).
\bibitem{Marandet2003} Y. Marandet \textit{et al.}, Communications in non linear
science and numerical simulations, \textbf{8}, 469 (2003).



\end{thebibliography}
\end{document}